\documentstyle[aps,multicol,psfig]{revtex} 
\newcommand{\ket}[1]{\left|#1\right\rangle}

\newcommand{\brkt}[3]{\left\langle#1\right|#2\left|#3\right\rangle} 
 
\newcommand{\proj}[2]{|#1\rangle\!\langle#2|} 
\newcommand{\vek}[1]{\mbox{\boldmath$#1$}} 
\newcommand{\equer} {\overline{\epsilon}} 
 
\newcommand{\Eq}[1]{Eq.(\ref{#1})} 
 
\newcommand{\punkt}[1]{\stackrel{\mbox{\boldmath$\cdot$}}{#1}} 
\newcommand{\dn}{d_{ng}} 
\newcommand{\darg}[1]{d_{#1 g}} 
\newcommand{\de}{d_{\epsilon g}} 
\newcommand{\stm}[1]{\left\langle#1\right\rangle}

\begin{document}   
\tighten   
\draft  
\author{B.Eggers$^*$ and G. Alber$^\dagger$} 
\address{$^*$Theoretische Quantendynamik, Universit\"at Freiburg, D--79104 Freiburg, 
\\$^\dagger$Abteilung f\"ur Quantenphysik, Universit\"at Ulm, D--89069 Ulm\\ 
(\underline{submitted to Phys. Rev. A. (February 2000))}} 
\title{\bf Stochastic dynamics of electronic wave packets in
fluctuating laser fields}
\maketitle 
\begin{abstract} 
The dynamics of a laser-excited 
Rydberg electron under the influence 
of a fluctuating laser field are investigated. 
Rate equations are developed which 
describe these dynamics 
in the limit of large laser bandwidths  
for arbitrary types of laser fluctuations. 
These equations apply whenever all coherent effects 
have already been damped out. 
The range of validity of these rate equations is 
investigated in detail 
for the case of phase fluctuations. 
The resulting asymptotic power laws 
are investigated which 
characterize the long time dynamics of the laser-excited 
Rydberg electron 
and it is shown to which extent these power laws 
depend on details of the laser spectrum. 
\end{abstract} 
\draft\pacs{42.50.Ct,32.50.Ar,42.50.Hz} 
 
\begin{multicols}{2} 
\section{Introduction}

The advancement of sophisticated trapping techniques and the 
development of powerful laser sources has stimulated numerous 
theoretical and experimental investigations on the 
dynamics of wave packets in elementary quantum systems. 
An understanding of their dynamics  
is important for our conception of quantum mechanics and 
its relation to classical mechanics. 
So far most of the research in this context has concentrated 
on coherent aspects of wave packet dynamics which may be traced back 
theoretically to semiclassical aspects originating from 
the smallness of the de Broglie wave lengths involved. 
However, for an understanding of the emergence of classical 
behaviour also a detailed understanding of the destruction 
of quantum coherence is required. Such a destruction of 
coherence may arise from the coupling of a quantum system 
to a reservoir or from stochastic external influences. 
Though by now many aspects  of the coherent dynamics 
of wave packets are well understood  still many questions 
concerning the influence of stochastic perturbations  
on elementary quantum systems with a high level density are open. 
 
A paradigm of a quantum system 
in which many of these latter aspects can be investigated in  
great detail are Rydberg systems interacting with fluctuating laser 
fields. Due to the inherent stochastic nature of laser light  
a detailed understanding of optical excitation processes 
with fluctuating laser fields is of vital interest for 
laser spectroscopy. 
Rydberg systems \cite{Seaton,Greene,Rau} 
are of particular interest in this context 
due to their high level density close to an ionization threshold. 
Any laser-induced excitation process which involves 
Rydberg and continuum states close to an ionization threshold 
typically leads to the preparation of a spatially localized 
electronic Rydberg wave packet \cite{AlberZoller}. 
Under the influence of a fluctuating laser field the coherence of 
such an electronic 
wave packet is disturbed whenever it is close to the ionic core 
where the electron-laser interaction is localized \cite{localized}. 
Eventually these random perturbations are expected to lead to 
a stochastically dominated Brownian motion of the electronic wave packet. 
Indeed it has been demonstrated recently \cite{Wir} that such a transition 
to a diffusive behaviour takes place  and that this diffusive 
dynamics are dominated by characteristic power laws which govern 
the time evolution of the Rydberg system. 
However, these previous investigations were restricted to a particular 
type of phase fluctuations of laser fields 
which can be described by the so called 
phase diffusion model (PDM) \cite{Haken,Scully}. This PDM 
implies a Lorentzian spectrum for the fluctuating laser field. 
It is known from the dynamics of three level systems 
that the somewhat unrealistic asymptotic frequency dependence 
of a Lorentzian laser spectrum may lead to unphysical predictions \cite{Dixit}. 
However, from our previous investigations it remained open to which extent 
these characteristic, 
diffusive long time dynamics of an excited Rydberg electron 
depend on details of the fluctuations of the exciting laser field.  
Such a dependence might be expected on intuitive grounds 
as the diffusion of the excited Rydberg electron in energy space 
eventually also reaches the far-off resonant regions of the laser spectrum. 
 
In this paper we tackle these open questions by 
generalizing our previous results to arbitrary types of 
laser fluctuations. For this purpose we derive rate equations 
for the relevant density matrix elements of the excited 
Rydberg electron which are averaged 
over the fluctuations of the laser field. 
These rate equations are based on a decorrelation of 
the relevant electron-field averages. 
This decorrelation approximation (DCA) is valid as long as the 
characteristic correlation time of the fluctuating laser field, 
i.e. its inverse bandwidth, is much smaller than 
all other relevant intrinsic dynamical time scales. 
Within this framework it will become apparent that it is the laser 
spectrum only which determines the time evolution of the 
excited Rydberg electron. 
On the basis of this approach it 
will be demonstrated which aspects of the diffusive long  
time dynamics of an excited electronic Rydberg wave packet 
depend on which details of the laser spectrum.  
 
The range of validity of these Pauli-type rate equations 
is investigated in detail for 
a special class of phase fluctuations \cite{Dixit,Elliot} 
of the exciting laser field. 
This special class of phase fluctuations 
implies non-Lorentzian spectra which for large 
laser frequencies decrease more rapidly 
than a Lorentzian. These phase fluctuations might be considered 
as a realistic model for  
a single mode laser field which is operated well above the laser threshold. 
In order to access the range of validity of the Pauli-type 
rate equations for this special class of laser fluctuations 
a more general master equation is derived 
for the averaged density operator of the Rydberg electron. 
This more general approach is also capable 
of dealing with all coherent aspects of the laser excitation process. 
 
The paper is organized in the following way: 
In Sec. II the theoretical models for describing the laser fluctuations 
and the dynamics of the Rydberg electron are presented. 
For the sake of simplicity we restrict our subsequent discussion 
to Rydberg systems which can be described within the framework 
of a one-channel approximation \cite{Seaton,Greene}. 
Typically this approximation is well satisfied for Alkali atoms. 
In Sec.III we derive rate equations which describe the dynamics of 
the excited Rydberg electron averaged over the laser fluctuations. 
Self consistent validity conditions for the applicability 
of the decorrelation approximation (DCA) 
are discussed on which these rate equations are based on. 
In Sec. IV characteristic aspects of the time evolution predicted 
by the rate equations of Sec. III are exemplified. The different 
dynamical long time regimes and their characteristic power laws 
are discussed in detail. From the resulting analytical expressions 
for these power laws it is apparent to which extent they 
depend on details of the laser spectrum. 
In Sec.V we derive a more sophisticated master equation 
for the averaged dynamics of the excited Rydberg electron. 
This master equation is capable of describing also coherent 
aspects of the dynamics of the excited Rydberg electron but 
its validity is restricted 
to a particular class of phase fluctuations only. 
In Sec. VI solutions of this master equation are compared 
with the corresponding results 
of the rate equations. Thus we are able to determine the range 
of validity of the DCA.  
In our subsequent discussions we use 
atomic units ($m_e=e=\hbar=1$). 
 
\section{Theoretical framework}\label{hamiltonsection} 
In this section the theoretical models are introduced with which 
the fluctuating laser field and the excited Rydberg system are described. 
\subsection{The fluctuating laser field} 
We consider an atomic or molecular Rydberg system  
which is driven by a laser field with electric field  
strength 
\begin{equation}\label{efeld} 
\vek{E}(t)=\vek{e}\varepsilon(t)e^{-i\omega t}+\mbox{c.c..} 
\end{equation} 
We assume that this laser field can be described by a classical 
stochastic process \cite{Haken}. 
The mean frequency of this laser field is denoted $\omega$ 
and $\vek{e}$ is its polarization vector. 
The fluctuations of this laser field are described by the 
envelope function $\varepsilon(t)$ which 
is assumed to be slowly varying on time scales of the order of 
$1/\omega$. 
The associated spectrum of this laser field is defined by \cite{Scully} 
\begin{equation}\label{Spektrum} 
S(\Omega)=\frac{1}{\pi}\mbox{Re}\int_0^\infty d\tau K(\tau)e^{-i\Omega\tau} 
\end{equation} 
with the two-time correlation function 
of the slowly varying amplitude 
\begin{equation}\label{allgemeinkorrelation} 
K(\tau)=\stm{\varepsilon(t+\tau)\varepsilon^*(t)}. 
\end{equation}  
Thereby 
$\langle...\rangle$ denotes statistical 
averaging over the fluctuations of the laser field. 
 
\begin{figure} 
\psfig{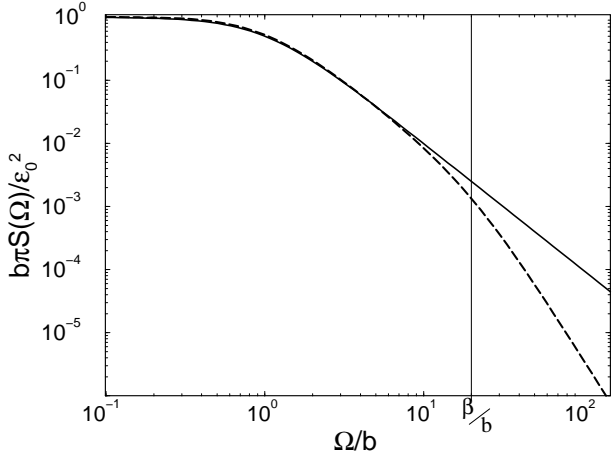} 
\begin{minipage}[t]{8.0 cm} 
\caption[]{Lorentzian laser spectrum according to \Eq{pdmspektrum} (solid line) and non-Lorentzian spectrum
with $\beta=20b$ according
to Eqs.(\ref{Spektrum}) and (\ref{Korrelation}) (dashed line).}
\label{spektrumsbild} 
\end{minipage} 
\end{figure}

For a single mode laser which is operated 
well above laser threshold  
to a good degree of approximation  
the amplitude of the laser field is stable.  
In these cases fluctuations of a realistic laser field 
can be described by a classical electromagnetic field 
whose phase is fluctuating, i.e. 
\begin{equation}\label{flupha} 
\varepsilon(t)=\varepsilon_0e^{-i\Phi(t)}. 
\end{equation} 
The  
fluctuating phase $\Phi(t)$  
obeys the (Ito-)stochastic differential equation \cite{Haken} 
\begin{equation}\label{Langevin} 
d\phi(t)=-\phi(t)\beta dt+\sqrt{2b}\beta dW(t) 
\end{equation} 
with $\phi(t) = \dot{\Phi}(t)$. 
In Eq.(\ref{Langevin}) $1/\beta$ determines 
the correlation time of the stochastic frequency 
$\phi(t)$ and $dW(t)$ is 
the differential of a real-valued Wiener process with zero mean 
and unit variance, i.e. 
$\stm{dW(t)}=0$, $\stm{dW(t)^2}=dt$ \cite{Kloeden}. 
Eqs.(\ref{flupha}) and (\ref{Langevin})  
imply the relation 
\begin{equation}\label{Korrelation} 
K(\tau)=|\varepsilon_0|^2\exp\left[-b\tau+b/\beta\left(1-e^{-\tau\beta}\right)\right]. 
\end{equation}  
In the limit of large values of $\beta$ 
$\Phi(t)$ itself approaches a real-valued Wiener process, i.e. 
\begin{equation}\label{PDM} 
d\Phi(t)=\sqrt{2b}dW(t). 
\end{equation}  
This limiting case constitutes 
the so called phase diffusion model (PDM). It implies 
a Lorentzian laser spectrum of the form 
\begin{equation}\label{pdmspektrum} 
S(\Omega)=|\varepsilon_0|^2\frac{1}{\pi}\frac{b}{b^2+\Omega^2}. 
\end{equation} 
For $\beta \gg b$ 
the parameter $\beta$ may be interpreted as a cut-off parameter 
of the laser spectrum. This becomes apparent by 
noting that for frequencies $\Omega\ll\beta$ 
the spectrum is always approximately Lorentzian 
whereas for large frequencies, i.e. $\Omega \gg \beta$, it tends 
to zero more rapidly as can be seen in Fig.\ref{spektrumsbild}. More precisely, 
for $\beta \gg b$ we obtain from Eqs.(\ref{Korrelation}) 
and (\ref{Spektrum}) the approximate relation 
\begin{eqnarray}\label{btkleinspektrum} 
S(\Omega)&=& 
\frac{|\varepsilon_0|^2}{\pi} 
\frac{b}{\Omega^2 + b^2}\frac{1}{1+\left(\frac{\Omega}{\beta}\right)^2}. 
\end{eqnarray} 
\subsection{The interaction Hamiltonian} 
Let us assume that 
the considered atom or molecule is 
prepared initially in an energetically low lying bound 
state $|g\rangle$ with energy $\epsilon_g$. Furthermore, 
it is excited resonantly by the fluctuating 
laser field to Rydberg and/or continuum states $|n\rangle$ 
with energies $\epsilon_n$ close 
to an ionization threshold. 
In the dipole and rotating wave approximation 
this excitation process can be described by the Hamiltonian  
\begin{equation}\label{ham2} 
H=\sum_{j=g,n}\proj{j}{j}\epsilon_j 
-\sum_n \left[\varepsilon(t) e^{-i\omega t}\dn\proj{n}{g} 
+\mbox{h.c.}\right]. 
\end{equation} 
The dipole matrix elements between the initial state 
$|g\rangle$ and the excited Rydberg states $|n\rangle$ are denoted 
$\dn=\brkt{n}{\vek{d\cdot e}}{g}$.  
It is understood that the sum over the excited states $|n\rangle$ appearing 
in Eq.(\ref{ham2}) includes also an integration over the adjacent continuum states. 
The energies of the excited Rydberg states and the energy dependence of the 
dipole matrix elements entering the Hamiltonian of Eq.(\ref{ham2}) 
can be determined with the help of quantum defect theory (QDT) \cite{Seaton}. 
In the case of excited Rydberg states which can be described within 
the framework of the one-channel approximation we find, for example, 
\begin{eqnarray}  
\epsilon_n &=& -\frac{1}{2(n - \alpha)^2},\nonumber\\ 
\dn &=& \de (n - \alpha)^{-3/2}\equiv \de |\epsilon_n - \epsilon_{n+1}|^{1/2}. 
\label{onechannel} 
\end{eqnarray}  
Thereby $\alpha$ denotes the quantum defect of the excited Rydberg series 
and $\de$ is the dipole matrix element between the initial state $|g\rangle$ 
and an energy normalized 
continuum state $|\epsilon \rangle$ with energy $\epsilon \approx 0$. 
Within the framework of QDT 
$\alpha$ as well as $\de$ are approximately energy independent close to the 
ionization threshold, i.e. for energies $|\epsilon|\ll 1$.  
A one channel approximation is appropriate for all cases in which excited states 
of the ionic core are located far away from the excited energy region. 
Typically this condition is fulfilled for Alkali atoms. 
 
The main problem is to solve the stochastic Schr\"odinger equation 
associated with the Hamiltonian of Eq.(\ref{ham2}). 
In general this is a complicated 
task due to the simultaneous presence of the laser fluctuations and 
of the threshold effects arising from the infinitely 
many bound Rydberg states converging to the ionization threshold. 
By the resulting intricate interplay between laser fluctuations  
and threshold phenomena it is difficult to apply stochastic simulation 
methods which typically become 
unreliable due to numerical inaccuracies in particular for long 
interaction times.

\section{Decorrelation approximation (DCA)}\label{DCAsection} 
 
In this section an approximation method is developed for 
determining the dynamics of the Rydberg system 
in the fluctuating laser field. 
This approximation method is based on a decorrelation of atom-field 
averages and leads to a Pauli-type master equation  
for the density operator of the Rydberg system which is averaged 
over the laser fluctuations. 
In this master equation all coherences between different energy levels 
have been eliminated adiabatically. This decorrelation approximation (DCA) 
is valid for arbitrary types of laser fluctuations provided 
these fluctuations are sufficiently fast (compare with conditions 
(\ref{dca1}), (\ref{dca2}) and (\ref{dcacont}) derived below). 
 
Let us start by determining first of all 
an approximate equation of motion 
for the probabilities $\rho_{nn}(t)=\langle n|\psi (t)\rangle \langle \psi(t)| n\rangle$ 
of observing the excited Rydberg system in one of the Rydberg states $|n\rangle$. 
Neglecting coherences $\rho_{nn'}(t)$ with $n\not=n'$ 
we find 
from the stochastic Schr\"odinger equation with Hamiltonian (\ref{ham2}) the relation 
\begin{eqnarray}\label{ib1} 
\punkt{\rho}_{nn}(t) &=& 2\mid \dn\mid^2 {\rm Re}\{ 
\int_{t_0}^t dt'\varepsilon(t)\varepsilon^*(t') 
e^{-i(\equer - \epsilon_n)(t - t')}\} 
\nonumber\\ 
&&\times\left[\rho_{gg}(t')-\rho_{nn}(t')\right] 
\nonumber\\ 
&&-\left[i\dn \varepsilon(t) e^{-i(\equer - \epsilon_n)(t - t_0)}  
\rho_{gn}(t_0)+ \mbox{h.c.}\right] 
\end{eqnarray} 
with $0\leq t_0 \leq t$ and with $\rho_{kj}(t) =  
\langle k|\psi(t)\rangle \langle \psi(t)|j\rangle
\exp[it(\epsilon_k-\epsilon_j)]$. 
The mean excited energy is denoted 
$\equer=\epsilon_g+\omega+\delta\omega$ with the  
quadratic Stark shift contribution of all other (non-resonant) 
states. In the subsequent discussion we assume for the sake 
of simplicity that the intensity dependence of 
$\delta\omega$ does not affect the dynamics of the excited 
Rydberg system. This is valid either for pure 
phase fluctuations of the laser field or in the case of 
arbitrary laser fluctuations for laser bandwidths 
which are much larger than $\delta\omega$. 
The  self-consistency condition for the omission of the coherences 
$\rho_{nn'}(t)$ with $n\not=n'$ 
will be discussed later (compare with Eq.(\ref{dca2})). 
Taking the integration interval $[t_0,t]$ to be  
smaller than the characteristic time scale 
over which $[\rho_{nn}(t')-\rho_{gg}(t')]$ changes significantly 
this latter term can be approximated by its value at time $t$. 
If on the other hand the interval 
$[t_0,t]$ is assumed to be larger than the correlation time of the fluctuating laser 
field  
we can replace the lower integration limit $t_0$ by $-\infty$ in Eq.(\ref{ib1}). 
Thus we obtain 
\begin{eqnarray}\label{ib3} 
\punkt{\rho}_{nn}(t)&=&2|\dn|^2\left[\rho_{gg}(t)-\rho_{nn}(t)\right]\\ 
&&\times\mbox{Re}\int_0^\infty d\tau\varepsilon(t)\varepsilon^*(t-\tau) 
e^{-i\tau(\equer-\epsilon_n)}\nonumber\\ 
&&-\Bigl\{i\dn\varepsilon(t)e^{-i(t-t_0)(\equer-\epsilon_n)} 
\rho_{gn}(t_0)+\mbox{h.c.}\Bigr\}.\nonumber 
\end{eqnarray} 
Now we are able to carry out the statistic average $\stm{...}$ over the 
laser fluctuations. Due to the above  
mentioned conditions on the integration interval the involved density  
matrix elements  
and the laser field $\varepsilon(t)$ decorrelate.  
As $\stm{\varepsilon(t)}=0$ 
the contribution of the last term on the right hand 
side of \Eq{ib3} vanishes. 
The remaining terms yield the rate equation 
\begin{equation}\label{raten1} 
\langle\punkt{\rho}_{nn}(t)\rangle ={\cal R}_{ng}\Bigl[\stm{\rho_{gg}(t)} 
-\stm{\rho_{nn}(t)}\Bigr] 
\end{equation}  
with the time independent rates 
\begin{equation}\label{rate} 
{\cal R}_{ng}=2\pi|\dn|^2S(\equer - \epsilon_n) 
\end{equation} 
and with 
the laser spectrum $S(\Omega)$ 
as defined by Eq.(\ref{Spektrum}). 
 
In order to work out quantitative criteria for the validity of  \Eq{raten1} 
let us define an effective 
bandwidth ${\cal B}$ of the fluctuating 
laser field $S(\Omega)$ by the relation 
\begin{equation}\label{Bdef} 
{\cal B}S(0)\pi\equiv\int_{-\infty}^\infty d\Omega S(\Omega)=\stm{|\varepsilon|^2}. 
\end{equation} 
The quantity $1/{\cal B}$ measures the correlation time of the fluctuating 
laser field. 
In the case of the PDM, for example, this effective bandwidth equals the width 
of the Lorentzian spectrum 
$b$. In all other cases which are described by Eq.(\ref{Langevin}) 
it characterizes the effective frequency width of the laser spectrum of 
Eq.(\ref{Spektrum}). 
According to Eq.(\ref{raten1}) the inverse rates $1/{\cal R}_{ng}$ 
define the characteristic time scale over which 
$[\rho_{gg}(t)-\rho_{nn}(t)]$ varies significantly. 
Thus, the above decorrelation approximation applies to cases only 
for which ${\cal B}\gg{\cal R}_{ng}$. 
Typically one finds ${\cal R}_{n_{res}g}\ge{\cal R}_{ng}$  
with $\epsilon_{n_{res}}=\overline{\epsilon}$ 
so that one of the validity conditions for the decorrelation condition becomes 
\begin{equation}\label{dca1a} 
{\cal B}\gg 
2\pi|\darg{n_{res}}|^2S(0). 
\end{equation} 
Eliminating $S(0)$ by \Eq{Bdef} we finally arrive at 
the equivalent condition 
\begin{equation}\label{dca1} 
{\cal B}^2\gg \frac{1}{2} 
\stm{\Omega_R}^2 
\end{equation} 
with the average Rabi-frequency 
$\stm{\Omega_R}=2|\darg{n_{res}}|\sqrt{\stm{|\varepsilon|^2}}$. 
What remains to be found is a validity condition  for neglecting the coherences  
$\rho_{nn'}(t)$ with $n\neq n'$ in \Eq{ib1}.  
As long as 
\begin{equation}\label{dca2} 
|\epsilon_{n_{res}}-\epsilon_{n_{res}+1}|\gg {\cal R}_{n_{res}g}= 
\frac{\stm{\Omega_R}^2}{2\cal{B}} 
\end{equation}  
the coherences $\rho_{nn'}(t)$ with $n\neq n'$ are 
rapidly oscillating functions in comparison with 
the slowly varying probabilities 
$\langle \rho_{nn}(t)\rangle $ and $\langle \rho_{gg}(t)\rangle $ 
entering Eq.(\ref{ib1}) 
so that their influence averages to zero approximately. 
Therefore the inequalities (\ref{dca1}) and (\ref{dca2}) are the  
required conditions for the validity of the DCA. 
According to these conditions we may distinguish two limiting cases. 
In the limit of small laser bandwidths for which 
${\cal B} \ll |\epsilon_{n_{res}}-\epsilon_{n_{res}+1}|$ 
they 
reduce to the requirement 
${\cal B}^2 \gg \stm{\Omega_R}^2/2$. In two-level systems 
which are excited resonantly by a fluctuating laser field 
this is the well known limit of large laser bandwidths in which 
the dynamics are dominated by rate equations. 
In the opposite limit where the bandwidth is large enough to 
affect many excited Rydberg states, i.e. 
${\cal B} \gg |\epsilon_{n_{res}}-\epsilon_{n_{res}+1}|$, 
the conditions for the applicability of the DCA reduce to 
the relation $\cal{B}\gg \gamma/\pi$. 
Thereby we have 
introduced the laser-induced 
rate  
\begin{equation} 
\gamma = 2\pi|\de |^2 \stm{|\varepsilon|^2}\equiv \frac\pi2 \stm{\Omega_R}^2 
|\epsilon_{n_{res}}-\epsilon_{n_{res}+1}|^{-1} 
\label{ionrate} 
\end{equation} 
which characterizes ionization 
of the initial state $|g\rangle$ into continuum states 
close to the ionization threshold according to Fermi's Golden rule. 
 
Up to now, our arguments for the 
derivation of the rate equation (\ref{raten1})  
and of conditions (\ref{dca1}) and (\ref{dca2})  
apply for discrete excited states only.  
However, our previous arguments can 
be generalized easily also to continuum states 
by viewing these continuum states as  
infinitesimally spaced discrete energy levels.  
According to quantum defect theory  
close to an ionization threshold 
the energy dependence of  
the discrete dipole matrix elements $\dn$ 
is described by Eq.(\ref{onechannel}). 
Thus  
condition (\ref{dca2})  
reduces to 
\begin{equation}\label{dcacont} 
{\cal B}\gg2|\de|^2\stm{|\varepsilon|^2}\equiv \frac{\gamma}{\pi}. 
\end{equation} 
In the limit of an infinitesimally small level spacing 
between the excited states 
Eqs. (\ref{raten1}) and (\ref{onechannel}) 
imply 
that the probability of finding the excited Rydberg system in  
a continuum state becomes vanishingly small. Thus we find  
\begin{equation} 
\langle \dot{\rho}_{\epsilon \epsilon} (t)\rangle  = {\cal R}_{\epsilon g} 
\langle \rho_{gg}(t)\rangle 
\end{equation} 
with ${\cal R}_{\epsilon g} = 2\pi |\de|^2 S(\overline{\epsilon} - \epsilon)$. 
Integration over the whole electron continuum finally yields 
\begin{equation}\label{raten2} 
\langle \punkt{P}_{ion}(t)\rangle =\Gamma \langle \rho_{gg}(t)\rangle 
\end{equation} 
with the mean ionization probability 
$\langle P_{ion}(t)\rangle =\int_0^{\infty} d\epsilon 
\langle \rho_{\epsilon\epsilon}(t)\rangle$  
and with the effective ionization rate 
\begin{equation} 
\Gamma = 2\pi \int_0^{\infty} d\epsilon |\de|^2 
S(\overline{\epsilon} - \epsilon). 
\label{effion} 
\end{equation} 
If the mean excited energy $\overline{\epsilon}$ 
is located well above threshold and if $\de$ is still energy 
independent over the energy region over which 
$S(\overline{\epsilon} - \epsilon)$ 
is significant, this effective ionization $\Gamma$ 
reduces to the previously introduced 
ionization rate $\gamma$ of Eq.(\ref{ionrate}). 
 
The rate Eqs. (\ref{raten1}) and (\ref{raten2}) together with  
the conservation of probability, i.e. 
\begin{equation} 
\langle \rho_{gg}(t)\rangle  = 1 - \langle P_{ion}(t)\rangle  - 
\sum_n \langle \rho_{nn}(t)\rangle 
\label{conservation} 
\end{equation} 
and together with the initial condition $\rho(t=0)=\proj{g}{g}$ 
determine the time evolution of a laser excited Rydberg electron within 
the framework of the DCA. 

\section{Stochastic dynamics of Rydberg systems within the DCA}\label{DCAAlkali} 
 
In this section the dynamics of a Rydberg system is investigated  
with the help of the DCA on the basis of Eqs.(\ref{raten1}), (\ref{raten2}) 
and (\ref{conservation}). 
 
Within the framework of a one-channel approximation the excited energies and 
the energy dependence of the relevant dipole 
matrix elements of a Rydberg system  
can be described by Eqs.(\ref{onechannel}). They are characterized by a  
quantum defect $\alpha$ and by an energy-normalized dipole matrix element 
$\de$ which are both approximately energy independent for $|\epsilon| \ll 1$. 
The laser-induced coupling between the initial state $|g\rangle$ and the excited 
Rydberg- and continuum states 
is characterized by the ionization rate $\gamma$ of 
Eq.(\ref{ionrate}). Typically this description is adequate for Rydberg 
states of Alkali atoms. 
 
The rate equations for the averaged density operator of the Rydberg system 
(compare with Eqs.(\ref{raten1}),(\ref{raten2}) and (\ref{conservation})) 
can be analyzed in a convenient way with the help of Laplace transformations. 
Defining the Laplace transformed density operator by 
\begin{equation}\label{trafo} 
\langle \tilde{\rho}(z)\rangle  = \int_0^{\infty} dt e^{i z t}\langle \rho(t)\rangle 
\end{equation} 
the associated inverse transformation is given by 
\begin{equation}\label{ruecktrafo} 
\langle \rho(t)\rangle \equiv\frac{1}{2\pi}\int_{-\infty+i0}^{\infty+i0} 
dz e^{-izt}\langle \tilde{\rho}(z)\rangle. 
\end{equation} 
Thus the Laplace 
transformed rate equations (\ref{raten1}), (\ref{raten2}) 
and (\ref{conservation}) 
imply the relations 
\begin{eqnarray} 
\langle \tilde{\rho}_{gg}(z)\rangle 
&=&\frac{1}{\Gamma -iz\sigma(z)}\label{rggz1}\\ 
\mbox{and }\langle \tilde{P}_{ion}(z)\rangle 
&=&\frac{i\Gamma}{z\left[\Gamma-iz\sigma(z)\right]} 
\label{pionz1} 
\end{eqnarray} 
with  
\begin{equation}\label{sigmastoch} 
\sigma(z)=\sum_{n}\frac{{\cal R}_{ng}}{{\cal R}_{ng}-iz}. 
\end{equation} 
The rates ${\cal R}_{ng}$ entering Eq.(\ref{sigmastoch}) characterize 
the transitions between states $|g\rangle$ and $|n\rangle$ within the DCA and  
are defined by Eq.(\ref{rate}). 
In the derivation of Eqs.(\ref{rggz1}) and (\ref{pionz1}) 
$\langle \rho_{gg} (t)\rangle$
has been
neglected in comparison with 
$\sum_{n}\langle \rho_{nn} (t)\rangle$ and
$\langle P_{ion} (t)\rangle$
in Eq.(\ref{conservation}).

We may distinguish various dynamics regimes 
which are treated subsequently.  
\subsection{Asymptotic long time behaviour}\label{DCAlangzeit} 
 
The time evolution of the averaged density operator of the Rydberg system 
can be obtained from Eqs.(\ref{rggz1}) and (\ref{pionz1}) 
and from the inversion formula (\ref{ruecktrafo}). 
In general the time evolution will exhibit both exponential decays  
originating from poles of the Laplace transforms (\ref{rggz1}) and 
(\ref{pionz1}) 
in the complex $z$-plane and  
power law decays which originate from cut contributions 
starting from the branch point of $\sigma(z)$ at $z=0$. 
As the asymptotic long time behaviour will be dominated by power law 
decays we have to investigate the structure of the characteristic kernel $\sigma(z)$ 
around the branch point $z=0$ in more detail. 
From Eqs.(\ref{sigmastoch}) and (\ref{rate}) 
it follows that for $z\to0$ its main contributions  
arise from the infinitely 
many Rydberg states very close to the ionization threshold.  
Hence in the long time limit we  
may approximate $S(\equer - \epsilon_n)$ by $S(\equer)$  
in expression (\ref{rate}). Furthermore we may 
replace the sum over all Rydberg states in 
Eq.(\ref{sigmastoch}) by an integration. So finally 
in the limit $z\to 0$ we obtain the relation 
\begin{equation}\label{sigmastochlangzeit} 
\sigma(z)\to \frac{2\pi}{3\sqrt{3}}\left(\frac{i2\pi 
|\de|^2 S(\equer)}{z}\right)^{1/3} \hspace{0.2cm}(z\to 0). 
\end{equation} 
Inserting Eq.(\ref{sigmastochlangzeit}) into 
Eqs.(\ref{ruecktrafo}), (\ref{rggz1}) and (\ref{pionz1}) 
one obtains the 
asymptotic long time behaviour 
\begin{eqnarray} 
\langle \rho_{gg}(t)\rangle &=&\left(\frac{S(\equer)}{\stm{|\varepsilon|^2}}\right)^{1/3} 
\frac{\Gamma(\frac 5 3)}{3(\Gamma/\gamma)^2}(t\gamma)^{-5/3}\label{rgglangzeit}\\ 
\langle P_{ion}(t)\rangle 
 &=&1-\left(\frac{S(\equer)}{\stm{|\varepsilon|^2}}\right)^{1/3} 
\frac{\Gamma(\frac 2 3)}{3(\Gamma/\gamma)}(t\gamma)^{-2/3}\label{pionlangzeit} 
\end{eqnarray} 
with $\Gamma(x)=\int_0^{\infty}du\;e^{-u} u^{x-1}$ denoting the gamma function \cite{AS}. 
Eqs.(\ref{rgglangzeit}) and (\ref{pionlangzeit}) 
describe the time evolution of the mean initial state probability 
and of the mean ionization probability 
for sufficiently long interaction times. 
They are generalizations 
of our previous results of Ref. \cite{Wir} which were only valid for 
phase fluctuations of the PDM. 
Within the framework of the DCA these asymptotic laws 
are valid for arbitrary fluctuations 
of the laser field  provided ${\cal B}\gg \gamma/\pi$ (compare with 
Eq.(\ref{dcacont})). 
Obviously this asymptotic long time behaviour is independent  
of the quantum defect which characterizes the 
influence of the ionic core of the Rydberg system. 
Furthermore, the characteristic exponents of the long time behaviour 
are a peculiar property of the Coulomb problem and do not depend 
on details of the laser spectrum. 
However, the time independent pre-factors of these 
power laws depend on $S(\Omega)$ and  
on the effective ionization rate $\Gamma$ of Eqs.(\ref{Spektrum}) 
and (\ref{effion}).  
 
After which interaction time 
do we expect the asymptotic power laws of Eqs.(\ref{rgglangzeit}) 
and (\ref{pionlangzeit}) to become valid? 
As apparent from Eq.(\ref{pionlangzeit}) 
this diffusive long time dynamics finally leads to complete ionization 
of the Rydberg system. Thus it is reasonable to characterize the onset of this 
asymptotic long time dynamics by a stochastic ionization time $t_c$ which is 
defined by the condition  
$\langle P_{ion}(t_c)\rangle = 1/2$ which yields 
\begin{equation}\label{tc} 
t_c=\frac{1}{\gamma}\left[ 
\left\{\frac{\gamma \Gamma(\frac{2}{3})}{\Gamma}\right\}^3\frac{8}{27} 
\frac{S(\equer)}{\stm{|\varepsilon|^2}}\right]^{1/2}. 
\end{equation}

\subsection{Intermediate interaction times}  
 
In this section we deal with characteristic aspects of the 
dynamics described by Eqs.(\ref{raten1}) and (\ref{raten2}) 
in cases in which 
the interaction times are large enough 
so that the initial state $|g\rangle$ is depleted significantly 
but which are still much smaller than 
the stochastic ionization time $t_c$ of Eq.(\ref{tc}). 
For these interaction times we may distinguish two characteristic 
regimes depending on whether the excited states are close to the ionization 
threshold or above or whether they are located well below this threshold. 
 
\subsubsection{Excitation at or above threshold }\label{anderschwelle} 
 
In this dynamical regime the significantly excited states are located at the 
ionization threshold or above. 
This implies that 
all rates ${\cal R}_{ng}$ which describe the coupling between states 
$|g\rangle$ and $|n\rangle$ 
are small in comparison with the total ionization rate $\gamma$. 
Thus considering interaction times $t$ which  
are not very much larger than 
$1/\gamma$ implies that we may 
take $t{\cal R}_{ng}\ll1$ for all quantum numbers $n$. Hence, considering the formal solution
of \Eq{raten1}
\begin{equation}
\stm{\rho_{nn}(t)}={\cal R}_{ng}\int_0^t ds e^{-(t-s){\cal R}_{ng}}\stm{\rho_{gg}(s)}
\end{equation}
we may replace the exponential by unity.
Performing the summation over all Rydberg states we find
with the help of $\sum_n{\cal R}_{ng}\approx\gamma-\Gamma$
(the sum over all Rydberg states has been replaced by an integral)
\begin{equation}\label{iw}
\sum_n\langle\punkt{\rho}_{nn}(t)\rangle=(\gamma-\Gamma)\stm{\rho_{gg}(t)}-\gamma^2
\Lambda_{Sp}\int_0^t ds\stm{\rho_{gg}(s)}
\end{equation}
where we used the quantity
\begin{eqnarray}\label{lambdadef} 
\Lambda_{Sp}&=& 
\frac1{\stm{|\varepsilon|^2}^2}\int_{-\infty}^0 
d\epsilon_n (\frac{d\epsilon_n}{dn})S^2(\equer - \epsilon_n)\equiv \nonumber\\ 
&& 
\frac1{\stm{|\varepsilon|^2}^2}\int_0^\infty 
d\epsilon S^2(\equer + \epsilon)(2\epsilon)^{3/2} 
\end{eqnarray} 
which characterizes the spectral properties of the fluctuating laser 
field.

We take the time-derivative of \Eq{conservation}
and eliminate the ionization probability with \Eq{raten2}. 
Inserting \Eq{iw} in this equation we find an integro-differential
equation for $\stm{\rho_{gg}(t)}$, namely   
\begin{equation}\label{rggintegrodiff}
\gamma\stm{\rho_{gg}(t)}-\gamma^2\Lambda_{Sp}\int_0^t ds \stm{\rho_{gg}(s)}+\langle\punkt{\rho}_{gg}(t)\rangle=0.
\end{equation}
As $\Lambda_{Sp}\ll1$ one finally obtains the relations 
\begin{equation}\label{rgganderschwelle} 
\langle \rho_{gg}(t)\rangle =\frac1{1+\Lambda_{Sp}}\Bigl[\exp(-\gamma t)+\Lambda_{Sp}\exp(\gamma\Lambda_{Sp}t)\Bigr] 
\end{equation}  
and  
\begin{equation}\label{pionanderschwelle} 
\langle P_{ion}(t)\rangle 
=\frac{\Gamma}{\gamma(1+\Lambda_{Sp})}\Bigl[\exp(\gamma\Lambda_{Sp}t)-\exp(-\gamma t)\Bigr]. 
\end{equation} 
These equations even apply to interaction times $t < 1/\gamma$.
Note that consistent with our approximations the interaction
times always fulfill the inequality $\gamma \Lambda_{Sp}t\ll 1$.
In the special case of the PDM 
$\Lambda_{Sp}$ reduces to 
\begin{equation}\label{lambdapdm} 
\Lambda_{Sp}= 
\sqrt{\frac b {2\pi^2}}\mbox{Re}\Bigl\{\bigl(1-2i\frac{\equer}{b}\bigr)\sqrt{\frac{\equer}{b}-i}\Bigr\}. 
\end{equation} 
Very close to threshold, i.e. for $\overline{\epsilon}\to 0$, one obtains 
$\Lambda_{Sp}=\sqrt{b}/(2\pi)$ so that in this special case we obtain again our 
previous results of Ref. \cite{Wir}. 
 
\subsubsection{Excitation well below threshold}\label{unterderschwelle} 
 
If the fluctuating laser field excites Rydberg states well below 
the ionization threshold, i.e.  ${\cal B}\ll|\equer|$ and $\equer<0$,   
the considerations of Sec.\ref{anderschwelle} have to be modified. 
Since the interaction time is assumed 
to be smaller than the stochastic ionization 
time $t_c$ 
we can approximate the characteristics of the 
dominantly excited Rydberg states  
by 
\begin{eqnarray} 
\epsilon_n&\to&\equer+2\pi (n - n_{res})/T_{\equer},\label{aequi1}\\ 
\dn &\to&\de \sqrt{2\pi/T_{\equer}}.\label{aequi2} 
\end{eqnarray} 
Thereby 
$T_{\equer}\equiv 2\pi(-2\overline{\epsilon})^{-3/2}$ denotes 
the classical Kepler period of the mean excited Rydberg state of 
energy $\equer<0$. 
Whereas in the previous subsection the  
stochastic influence could be characterized by a single average spectral 
property for arbitrary types of fluctuations, namely by $\Lambda_{Sp}$, 
the excitation dynamics well below threshold turns out to be much more 
sensitive to the details of the laser spectrum. 
This is easily demonstrated by considering phase fluctuations 
which can be described by the laser spectrum of 
Eq.(\ref{btkleinspektrum}) as a particular example.  
This spectrum describes fluctuations  
of a single mode laser field well above the laser threshold  
in the limit $\beta \gg b$. 
In addition, if only Rydberg states well below threshold are excited significantly 
we may neglect the effective ionization rate $\Gamma$  
in the denominator of   
Eqs.(\ref{rggz1}) and (\ref{pionz1}). 
In the limit $\beta \gg b$ we thus arrive at the relations  
\begin{eqnarray} 
\langle \rho_{gg}(t)\rangle 
&=&\frac{2}{T_{\equer}\beta}f'\left(\frac{2\gamma b}{T_{\equer}\beta^2} t\right) 
\label{rggbtklein}\\ 
\langle P_{ion}(t)\rangle &=&\frac{1}{\pi} 
\left[\frac{\beta}{|\equer|}-\arctan\left(\frac{\beta}{|\equer|}\right)\right]f 
\left(\frac{2\gamma b}{T_{\equer}\beta^2} t\right)\label{pionbtklein}. 
\end{eqnarray} 
Thus within this limit for arbitrary values of $\beta$ and $b$ 
the influence of the phase fluctuations of the laser field 
is described by the single scaling 
function $f(\tau)$ which is defined by the equation 
\begin{equation}\label{fvontau} 
\frac{df(\tau)}{d\tau}=-\mbox{Im}\int_0^{\infty}\frac{d\zeta}{\zeta} 
e^{-i\zeta\tau}\Bigl\{\int_{-\infty}^{\infty} 
\frac{dx}{1-i\zeta(x^2+x^4)}\Bigr\}^{-1}. 
\end{equation} 
To end up with \Eq{fvontau} we had to apply the further approximation $\Omega\gg b$
in the laser spectrum of Eq.(\ref{btkleinspektrum}).  
Physically speaking this
approximation means that we consider 
cases in which the essential dynamics are dominated by energy states
which are located in the wings of the laser spectrum.

\begin{figure} 
\psfig{file=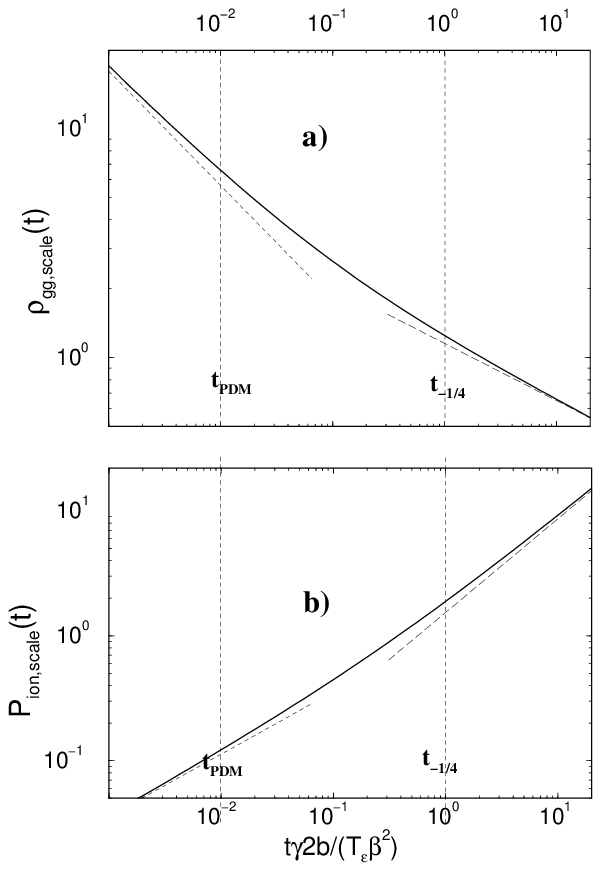,width=8.5cm}
\begin{minipage}[t]{8.0 cm} 
\caption[]{
Numerical solutions of Eqs.(\ref{rggbtklein}) and (\ref{pionbtklein})
\newline{\bf a) initial-state probability}
\newline $\rho_{gg,scale}(t)\equiv\stm{\rho_{gg}(t)}\frac12 T_{\equer}\beta$
(solid line) together with the asymptotic behaviour according to \Eq{alt1} (dashed) and \Eq{neu1} (long dashed).
\newline{\bf b) ionization probability}
\newline $P_{ion,scale}(t)\equiv
\pi[\beta/|\equer|-\arctan(\beta/|\equer|)]^{-1}\stm{P_{ion}(t)}$ 
(solid line) and asymptotic 
behaviour according to \Eq{alt2} (dashed) and \Eq{neu2} (long dashed).
}\label{fvonx} 
\end{minipage} 
\end{figure}

In the limits $\tau\ll 1$ and $\tau\gg 1$ asymptotic expressions are easily 
obtained from Eq.(\ref{fvontau}). 
The limit of small values of $\tau$ is realized in the PDM 
where $\beta \to\infty$ 
and where the spectrum of Eq.(\ref{btkleinspektrum}) 
reduces to a Lorentzian form. 
In this case one obtains the expression \cite{Wir} 
\begin{equation} 
f(\tau)\to 2\sqrt{\frac{\tau}{\pi}} \hspace{0.5cm}(\tau\ll 1). 
\end{equation} 
Consequently 
Eqs.(\ref{rggbtklein}) and (\ref{pionbtklein}) yield 
\begin{eqnarray} 
\langle \rho_{gg}(t)\rangle &=&\sqrt{\frac{2}{\pi bT_{\equer}\gamma t}}, 
\label{alt1}\\ 
\langle P_{ion}(t)\rangle &=&2 
\left[\frac 1 {|\equer|}-\frac{1}{\beta}\arctan\left(\frac{\beta}{|\equer|}\right)\right] 
\sqrt{\frac{2 t\gamma b}{\pi^3 T_{\equer}}}.\label{alt2} 
\end{eqnarray} 
From the numerical data shown in Fig.\ref{fvonx} it is apparent 
that Eqs.(\ref{alt1}) and (\ref{alt2})  
are good estimates for interaction times  
$t<t_{\mbox{{\scriptsize {PDM}}}}=T_{\equer}\beta^2/(200b\gamma)$.
In the extreme opposite limit of large values of 
$\tau$ we obtain the relations 
\begin{equation} 
f(\tau)\to\frac 4{3\pi}\Gamma(\frac 1 4)\tau^{3/4} \hspace{0.5cm}(\tau \gg 1) 
\end{equation} 
and 
\begin{eqnarray} 
&&\langle \rho_{gg}(t)\rangle 
=\frac{\Gamma(\frac{1}{4})}{\pi}\Bigl(\frac{8}{T^3_{\equer}\beta^2 t 
\gamma b}\Bigr)^{1/4},\label{neu1}\\ 
&&\langle P_{ion}(t)\rangle =\label{neu2}\\ 
&&\frac{4}{3\pi^2}\Gamma(1/4)\left[\frac \beta {|\equer|}- 
\arctan\left(\frac{\beta}{|\equer|}\right)\right] 
\Bigl(\frac{2\gamma b t}{T_{\equer}\beta^2}\Bigr)^{3/4}.\nonumber 
\end{eqnarray} 
In this case the power law decays which 
characterize the diffusive dynamics of the 
excited Rydberg electron differ from the 
corresponding results of the PDM significantly. 
Even the characteristic exponents 
are changed. 
According to Fig.\ref{fvonx} 
this dynamical regime is realized for interaction times $t$ which fulfill 
the relation 
$t_{-1/4}<t\ll t_c$ with 
$t_{-1/4}=T_{\equer}\beta^2/(2b\gamma)$. 

\section{Full master-equation}\label{FMaster} 
Due to their simplicity and their applicability to all types of laser spectra  
the  DCA rate equations are ideal for understanding 
the dynamics of Rydberg systems in the case of large laser bandwidths 
(compare with Eqs.(\ref{dca1}),(\ref{dca2}) and (\ref{dcacont})). 
However the DCA approximation is not capable of describing coherent 
aspects of the laser-induced excitation process. 
In order to investigate the limits of applicability of the DCA rate equations 
in this section a more general approach is developed which is also capable 
of describing all coherent aspects of the excitation process. 
For the sake of simplicity we shall restrict our subsequent discussion to the case 
of phase fluctuations of the exciting laser field which deviate only 
slightly from a Lorentzian spectrum and which can be modelled 
by Eqs.(\ref{flupha}), (\ref{Langevin}) and (\ref{btkleinspektrum}). 
For these type of laser fluctuations we shall derive an approximate master equation 
involving the density matrix elements of the 
excited Rydberg electron which are averaged over the fluctuations of the laser field. 
This procedure is a generalization of previous approaches which so far have been 
applied to atomic few level system only \cite{Dixit}. 
In the special case of the PDM these subsequently derived 
density matrix equations reduce to our 
previous results of Ref.\cite{Wir}. 
 
We start from the Schr\"odinger equation with Hamiltonian (\ref{ham2}) with 
a fluctuating laser field as given by Eqs. (\ref{flupha}) and (\ref{Langevin}). 
For the effective density operator 
\begin{equation}\label{rhotrafo} 
\rho(t)=\sum_{k,j\in{g,n}}|k\rangle \langle j| 
\langle k|\psi(t)\rangle \langle \psi(t)|j\rangle 
e^{i(\Phi(t)+\omega t)(\delta_{jg}-\delta_{kg})} 
\end{equation}   
we obtain the equation of motion 
\begin{equation}\label{Vonneumann2} 
\punkt{\rho}(t)=-i\Bigl[H_{dr.},\rho(t)\Bigr]-i\phi(t) 
\Bigl[\proj{g}{g},\rho(t)\Bigr]. 
\end{equation}  
The self-adjoint Hamiltonian 
\begin{equation}\label{Hdr} 
H_{dr.}=\sum_{n,g} \proj{n}{n}\epsilon_n+ 
\proj{g}{g}\equer-\varepsilon_0\sum_n\left(\dn\proj{n}{g}+\mbox{h.c.}\right) 
\end{equation} 
describes the dynamics of the Rydberg system in the absence of phase fluctuations 
and the stochastic process $\phi(t)$ is defined by Eq.(\ref{Langevin}). 
In order to average  
Eq.(\ref{Vonneumann2}) over all possible realizations of the stochastic process 
$\phi(t)$  
it is convenient to introduce the averaged operators 
\begin{equation}\label{rhondef} 
\rho^{(n)}(t)=\left(\frac{\beta}{b}\right)^{n/2}\frac{(-i)^n}{\sqrt{n!}} 
\int_{-\infty}^{\infty}d\phi\;Q_n(\phi)\rho(t)p(\phi,t) 
\end{equation}  
with $n=0,1,2,3,...$~. 
The (conditional) probability distribution $p(\phi,t)$  
obeys the Fokker-Planck equation 
\begin{equation} 
[\frac{\partial}{\partial t} + {\cal L}] p(\phi,t) = 0 
\end{equation} 
with the Fokker-Planck operator 
\begin{equation} 
{\cal L} = \beta \frac{\partial}{\partial \phi} \phi + 
b\beta \frac{\partial^2}{\partial \phi^2}. 
\end{equation} 
This equation has to be solved with the initial condition 
\begin{equation}\label{P0} 
p(\phi,0)=\frac 1 {\sqrt{2\beta b\pi}} \exp\left(-\frac{\phi^2}{2\beta b}\right) 
\end{equation} 
which represents the stationary solution of the Fokker-Planck equation. 
The quantities 
\begin{equation}\label{Qnloesung} 
Q_n(\phi)=H_n\Bigl(\phi\left[2^{n+1}n!\beta b \right]^{-1/2}\Bigr) 
\end{equation} 
with the Hermite polynomials $H_n$  
are eigenfunctions of the adjoined Fokker-Planck operator 
${\cal L}^{\dagger}$ with eigenvalues $\Lambda_n = n\beta$ \cite{AS}. 
Starting from Eq.(\ref{Vonneumann2}) 
we obtain 
a set of coupled differential equations  
for the operators $\rho^{(n)}(t)$, namely 
\begin{eqnarray}\label{Hierarchie} 
\punkt{\rho^{(n)}}(t)&=&-i\Bigl[H_{dr.},\rho^{(n)}(t)\Bigr]-n\beta \rho^{(n)}(t)\\ 
&+&b(n+1)\Bigl[\proj{g}{g},\rho^{(n+1)}(t)\Bigr] 
-\beta\Bigl[\proj{g}{g},\rho^{(n-1)}(t)\Bigr].\nonumber 
\end{eqnarray}    
These equations have to solved subject to the initial condition 
\begin{equation}\label{abkonkret} 
\rho^{(n)}(0)=\delta_{n0}\rho(0). 
\end{equation} 
According to Eqs.(\ref{rhondef}) and (\ref{Qnloesung}) 
$\rho^{(0)}(t)$ is the required density operator which is averaged over the phase 
fluctuations of the laser field. 
 
We may derive an approximate equation of 
motion for the averaged density operator 
$\rho^{(0)}(t)$. 
As a first step we Laplace transform Eqs.(\ref{Hierarchie}), i.e. 
\begin{eqnarray} 
-iz\tilde{\rho}^{(n)}(z)&=&\rho^{(n)}(0)-n\beta\tilde{\rho}^{(n)}(z)-\nonumber\\ 
&-&i\Bigl[H_{dr.},\tilde{\rho}^{(n)}(z)\Bigr]+\label{laplacehierarchie}\\ 
&+&\Bigl[\proj{g}{g},b(n+1)\tilde{\rho}^{(n+1)}(z)- 
\beta\tilde{\rho}^{(n-1)}(z)\Bigr].\nonumber 
\end{eqnarray}  
In view of the large values of $\beta$ we are interested in we may 
neglect terms containing $\varepsilon_0$  
in comparison with terms containing $\beta$. 
Thus  
with the definition  
\begin{equation}\label{adef} 
\tilde{\alpha}_l^n(z)=\frac{\tilde{\rho}^{n+1}_{lg}(z)}{\tilde{\rho}^{n}_{lg}(z)} 
\end{equation} 
 
we arrive at 
the recursion relations 
\begin{equation}\label{Rekursion} 
n+1+\frac{i}{\beta}(\epsilon_l-\equer-z)= 
-\frac{b}{\beta}(n+2)\tilde{\alpha}_l^{n+1}(z)+\frac{1} 
{\tilde{\alpha}_l^n(z)}. 
\end{equation} 
From these recursion relations we find 
\begin{equation}\label{Kettenbruch} 
\tilde{\alpha}_k^0(z)= 
\frac{1}{1+\displaystyle i\frac{\epsilon_k-\equer-z}{\beta}+\frac{2b/\beta} 
{2+\displaystyle i\frac{\epsilon_k-\equer-z}{\beta}+ 
\frac{3b/\beta}{3+\displaystyle ...}}}. 
\end{equation} 
Using Eqs.(\ref{adef})  and  
(\ref{Kettenbruch}) we may now eliminate 
$\tilde{\rho}^{(1)}(z)$ in Eq.(\ref{laplacehierarchie}) for $n=0$. 
Performing the Laplace back 
transformation (\ref{ruecktrafo}) 
and using  
the definition $\langle \rho(t)\rangle = 
\rho^{(0)}(t)$ 
we finally obtain the master equation 
\begin{eqnarray}\label{Master_gedaechtnis} 
\langle \punkt{\rho}(t)\rangle &=&-i\Bigl[H_{dr.},\langle \rho(t)\rangle\Bigr]\\ 
&-&b\int_0^t d\tau\sum_{k\not= g}\Bigl\{\alpha_k^0(\tau)\proj{k}{k} 
\langle\rho(t-\tau)\rangle\proj{g}{g} 
+\mbox{h.c.}\Bigr\}\nonumber 
\end{eqnarray} 
with the memory function 
\begin{equation}\label{alphaknull} 
\alpha_k^0(\tau)=\frac1{2\pi}\int_{-\infty+i0}^{\infty+i0}dze^{-iz\tau}\tilde{\alpha}_k^0(z). 
\end{equation}   
In the limit of the PDM, i.e. for 
$\beta\to\infty$, this master equation reduces to the well known form 
\cite{Agarval} 
\begin{eqnarray}
\langle\punkt{\rho}(t)\rangle 
&=&-i\Bigl[H_{dr.},\langle\rho(t)\rangle\Bigl]\label{PDMMaster}\\ 
&+&\frac 1 2\Bigl\{\bigl[L,\langle\rho(t)\rangle L^\dagger\bigr]+ 
\bigl[L\langle\rho(t)\rangle,L^\dagger\bigr]\Bigr\}\nonumber 
\end{eqnarray} 
with the Lindblad operator  
\begin{equation}\label{Lindbladoperator} 
L=\sqrt{2b}\proj{g}{g}. 
\end{equation} 
\section{Numerical results}\label{diskussion} 
 In this section 
numerical solutions of the master equation (\ref{Master_gedaechtnis}) 
are compared with the corresponding solutions of the DCA rate equations 
(\ref{raten1}),(\ref{raten2}) and (\ref{conservation}). 
Details of the numerical technique for solving Eq.(\ref{Master_gedaechtnis}) 
are summarized in the appendix. 
On the basis of these comparisons the validity conditions for the applicability 
of the DCA and its accuracy can be tested.  
For this purpose we consider the laser excitation of a Rydberg system which 
can be described by quantum defect theory in a one-channel approximation 
(compare with Eqs.(\ref{onechannel})). Typically this is a good approximation 
for Alkali atoms. 

\begin{figure} 
\psfig{file=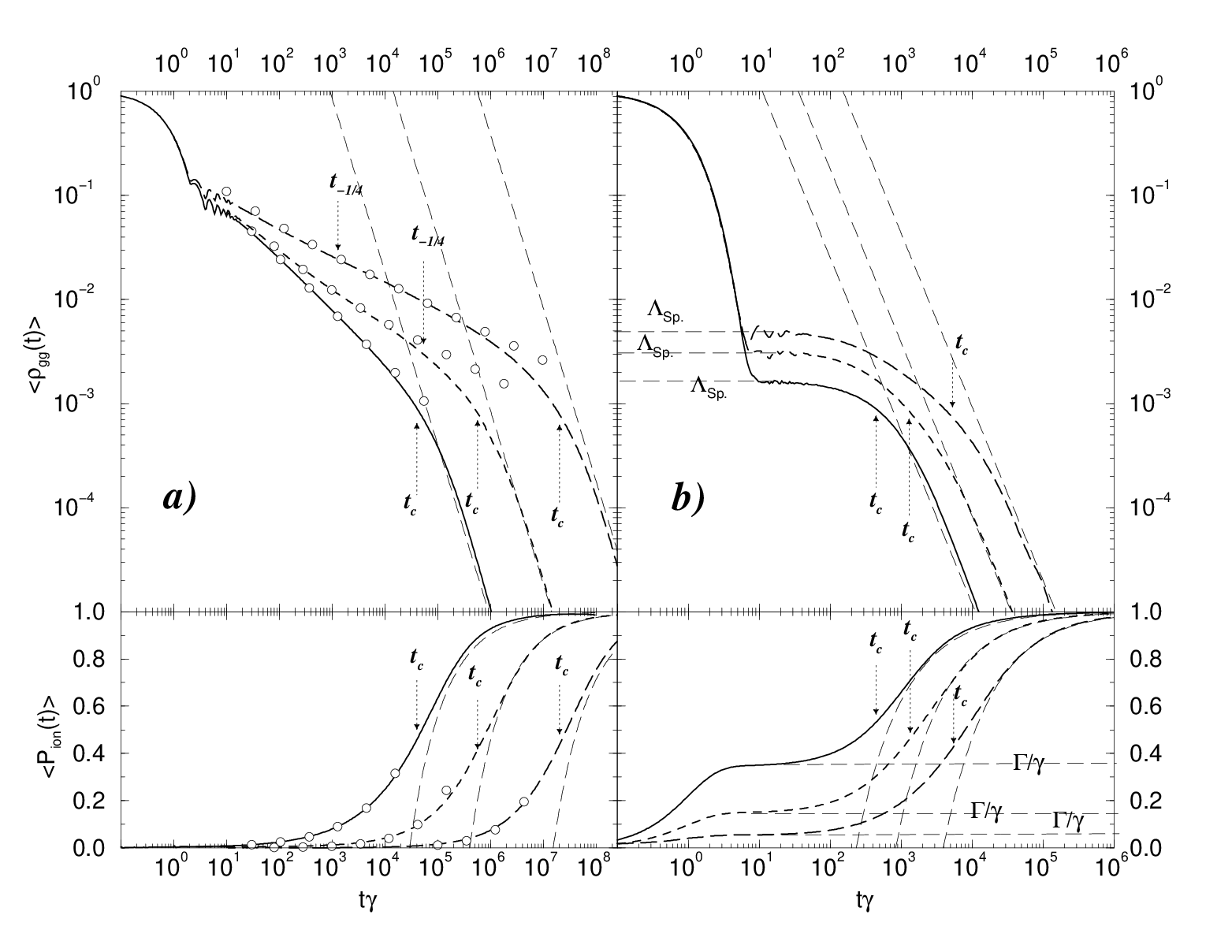,width=17cm,angle=270} 
\begin{minipage}[t]{17 cm} 
\caption[]{Mean initial state probability $\langle \rho_{gg}(t)\rangle$ 
and mean ionization probability $\langle P_{ion}(t)\rangle $ as a function of 
interaction time $t$ (in units of $1/\gamma$) for different values of $\beta$.   
\newline {\bf a) excitation well below threshold :} $b/\gamma=5,T_{\equer}\gamma=2, 
n_{res}=(-2\equer)^{-1/2}=200,\alpha=0.1$. PDM-limit 
 $b/\beta=0$: (solid curve), 
$b/\beta=0.03$: (dashed curve), $b/\beta=0.2$: (long dashed curve), 
homogeneously spaced energy level limit 
according to Eqs.(\ref{rggbtklein},\ref{pionbtklein}): 
(circles) and long time estimates according to 
Eqs.(\ref{rgglangzeit},\ref{pionlangzeit}): (thin long dashed 
curves)  
\newline {\bf b) excitation at threshold :} 
$b/\gamma=120,T_{\equer}\gamma=10, \equer/\gamma=-63.27,
(-2\equer)^{-1/2}=200,\alpha=0.1$. PDM limit  
($b/\beta=0$): (solid curve), strongly non-Lorentzian situations 
$b/\beta=3$: (dashed curve) and $b/\beta=10$: 
(long dashed curve). 
Long time estimates according to Eqs.(\ref{rgglangzeit},\ref{pionlangzeit}):
(thin long dashed curves).}\label{dynamik} 
\end{minipage} 
\end{figure}

The time evolution of the mean initial state probability $\langle \rho_{gg}(t)\rangle$ 
and of the mean ionization probability $\langle P_{ion}(t)\rangle$ are depicted in 
Fig.\ref{dynamik} for excitation at and well below the ionization threshold 
for different values of $\beta$. 
In both cases it is assumed that the exciting laser field has a well defined 
amplitude and a fluctuating phase.

Let us first turn to the case depicted in 
Fig.\ref{dynamik}a: 
The spectrum of the laser field is close to Lorentzian ($\beta\gg b$), so that the asymptotic form 
\Eq{btkleinspektrum} applies well. Thus the effective bandwidth $\cal B$ as defined by \Eq{Bdef} is 
approximately equal to the parameter $b$ which characterizes the spectrum of \Eq{btkleinspektrum} and the
parameter $\beta$ might be interpreted as an effective cut-off frequency of the laser spectrum.
 Rydberg states are excited by the fluctuating laser field 
well below the ionization threshold. 
The mean excited energy corresponds to a quantum number 
$n_{res}=(-2\equer)^{-1/2}=200$.  
The laser bandwidth $b$ and the laser-induced rate $\gamma$ are so small that 
the excited Rydberg states are located well below threshold, i.e. 
$-\equer\gg b,\gamma$. 
However, the values of $b$ and $\gamma$ are large enough so that more than 
one Rydberg state around energy $\overline{\epsilon}$ is affected significantly 
by the laser field, i.e. 
$T_{\overline{\epsilon}} \gamma, T_{\overline{\epsilon}} b > 1$. 
The three curves  of Fig.\ref{dynamik}a 
(solid, dashed and long dashed) correspond to different values of the 
effective cut-off frequency $\beta$ of the laser spectrum. 
As many excited states are involved in the depletion of state 
$|g\rangle$ the initial stage of the time evolution is governed by an approximate 
exponential decay of state $|g\rangle$ with rate $\gamma$ \cite{AlberZoller}. 
This initial stage of the time evolution is independent of the fluctuations of the laser field. 
At larger interaction times with  $t > 1/\gamma$ 
a coherent oscillation starts to appear in $\langle \rho_{gg}(t)\rangle$ 
with the classical Kepler period $T_{\overline{\epsilon}}$. 
This oscillation reflects the time evolution of the electronic Rydberg wave packet 
which has been prepared by the fast depletion of the initial state $|g\rangle$. 
With each return to the core region this Rydberg wave packet might undergo a transition 
to state $|g\rangle$ thus increasing $\langle \rho_{gg}(t)\rangle$. 
These coherent oscillations cannot be described by the DCA rate equations. 
However, due to laser fluctuations 
after a few Kepler periods these coherent oscillations are damped out 
and merge into diffusive dynamics which is characterized by power law decay of the 
initial state $|g\rangle$. 
From this time on the dynamics of the Rydberg system under the influence 
of the fluctuating laser field is well described by the DCA rate equations. 
This is apparent from Fig.\ref{dynamik}a by comparing the numerical solutions 
of the master equation (solid, dashed and long dashed curves) 
with the asymptotic solutions of the DCA rate equations 
(circles and thin dashed curves). 
According to the discussion presented in Sec. \ref{unterderschwelle} 
this diffusive dynamical behaviour appears when all coherent effects are damped out 
and disappears again at interaction times 
$t>t_c$ at which stochastic ionization starts to dominate. 
Physically speaking for these intermediate interaction times  
the excited electronic Rydberg wave packet starts to diffuse 
in energy space towards the ionization threshold. 
It reaches the ionization threshold roughly at time $t_c$ at which 
the ionization probability rises significantly from vanishingly small 
values to values close to unity. 
The early stages of this diffusion towards the ionization threshold 
are governed by the power law decay of Eq.(\ref{alt1}) for 
$\langle \rho_{gg}(t)\rangle$ which is characterized 
by the exponent $(-1/2)$. 
In the case of laser fluctuations which can be described by the PDM 
to a good degree of approximation 
this power law decay governs the time evolution 
up to the stochastic ionization time $t_c$. However, for non-Lorentzian 
spectra this is no longer the case. In cases in which the non-Lorentzian 
effects can be described by the spectrum of Eq.(\ref{btkleinspektrum}) 
the characteristic exponent of this power law decay is changed to 
a value of $(-1/4)$ 
as soon as the interaction times become larger than 
the characteristic time 
$t_{-1/4}=T_{\overline{\epsilon}}\beta^2/(2b\gamma)$ 
provided $t_{-1/4} < t_c$ (compare with Eq.(\ref{neu1})). 
This non-Lorentzian effect is clearly apparent in Fig.\ref{dynamik}a 
where the characteristic times $t_{-1/4}$ are indicated for 
$b/\beta=0.03$ and $b/\beta=0.2$. With increasing values of $\beta$ 
this characteristic time increases and this cross over phenomenon 
disappears for sufficiently large values of $\beta$ 
as soon as $t_{-1/4}>t_c$. 
At interaction times exceeding the stochastic ionization time 
$t_c$ the excited Rydberg wave packet has 
already reached the ionization 
threshold and the mean ionization probability rises to a value of 
unity. 
This asymptotic long time behaviour of the excitation dynamics 
is described to a good degree of approximation by 
Eqs.(\ref{rgglangzeit}) and (\ref{pionlangzeit}). 
That is apparent from a comparison of the thin dashed lines 
of Fig.\ref{dynamik}a with the corresponding numerical solutions 
of the master equation (\ref{Master_gedaechtnis}). 
\end{multicols} 
$\!\!\!\!$\hrulefill$\quad\hfill\ $\\ 
\begin{multicols}{2}   
In Fig.\ref{dynamik}b the laser bandwidth is so large that  
the significantly excited energy region $[\equer-{\cal B},\equer+{\cal B}]$ 
(before the onset of the electronic diffusion process) contains already 
the ionization threshold. 
Thus the excited Rydberg system is ionized  
significantly already in the early stages of the time evolution. 
As ${\cal B}\gg \gamma$  
this early stage of the ionization process is well described 
by the DCA rate equations which yield (compare to Eqs.(\ref{rgganderschwelle}) 
and (\ref{pionanderschwelle}))
\begin{eqnarray}
 \langle \rho_{gg}(t)\rangle &=& e^{-\gamma t}, \nonumber\\ 
 \langle P_{ion}(t)\rangle &=& \frac{\Gamma}{\gamma}(1 - e^{-\gamma t}). 
 \end{eqnarray} 
 These approximate solutions are obtained from Eqs.(\ref{raten1}) and 
 (\ref{raten2}) by neglecting 
$\langle \rho_{nn}(t)\rangle$ in comparison with $\langle \rho_{gg}(t)\rangle$. 
 According to 
 Eqs.(\ref{rgganderschwelle}) and (\ref{pionanderschwelle}) 
 this initial ionization process saturates  
 as soon as the  
 mean initial state probability and the mean ionization 
 probability have reached the values $\Lambda_{Sp}$ and $\Gamma/\gamma$. 
 At these interaction times  we still have $\gamma\Lambda_{Sp}t \ll 1$. 
 These characteristic aspects of the laser-induced excitation 
 process are clearly apparent in Fig.\ref{dynamik}b. 
 Physically speaking 
 in this initial stage of the excitation process 
 the Rydberg electron is ionized with probability $\Gamma/\gamma$. 
 With a probability of $(1- \Gamma/\gamma)$ 
 an excited electronic Rydberg wave packet is prepared after a time 
 of the order of $1/\gamma$. This wave packet is formed by a coherent 
 superposition of all Rydberg states within the dominantly excited 
 energy interval $[\equer-{\cal B},0]$. 
 Depending on the actual value of the laser bandwidth the coherent 
 dynamics of this electronic wave packet is damped sooner or later. After 
the destruction of all coherences to a good degree of approximation  
 the subsequent dynamics is governed by the DCA rate equations. 
 In Fig. \ref{dynamik}b small coherence oscillations are visible 
 in the time evolution of $\langle \rho_{gg}(t)\rangle$. 
 As soon as the interaction time exceeds the stochastic ionization time 
 the excited Rydberg electron starts to ionize significantly. The time 
 evolution of this stochastic ionization process is well described by  
 Eqs.(\ref{rgglangzeit}) and (\ref{pionlangzeit}) within the framework of the 
 DCA rate equations.    
 \section{Summary and conclusion} 
 The dynamics of an electronic Rydberg wave packet 
 under the influence of a fluctuating cw-laser field has been discussed. 
 It has been shown that for large laser bandwidths its dynamics 
 can be described by Pauli-type rate equations for the 
 relevant density matrix elements of the excited Rydberg electron 
 averaged over the laser fluctuations. These rate equations are 
 valid for arbitrary types of laser fluctuations and their dynamics 
 is determined by the spectrum of the laser field only and not by 
 any of the higher order correlation functions. 
 The validity of these rate equations has been investigated in detail 
 for a special class of phase fluctuations of the laser field. 
 
 With the help of these rate equations we have investigated the 
 dynamics of a laser excited Rydberg electron for long interaction times.  
 At these interaction times the dynamics of the Rydberg electron are 
 dominated by stochastic diffusion in energy 
 space towards the ionization threshold which leads finally to 
 stochastic ionization. This diffusion process is 
 accompanied by a characteristic scenario of power law decays. 
 Analytical expressions have been derived for these power laws and 
 their associated  characteristic exponents. 
 These analytical expressions exhibit in a clear way  
 that the asymptotic power laws 
 are independent of the quantum 
 defect of the excited Rydberg states and 
 to which extent they 
 depend on details of the laser spectrum. 
 In particular, it has been demonstrated that the characteristic exponents 
 which describe the process of stochastic ionization are completely 
 independent of the laser spectrum. 
 However, the initial stages of the diffusion of 
 the excited Rydberg electron depend on details of the laser spectrum.

 Support by the Deutsche Forschungsgemeinschaft within the SPP 
 `Zeitabh\"angige Ph\"anomene und Methoden' is acknowledged. 
 \end{multicols} 
 $\!\!\!\!$\hrulefill$\quad\hfill\ $\\ 
 
 \begin{appendix} 
\hfill \hrulefill\\ 
\begin{multicols}{2} 
\section{Numerical solution technique for the Master equation}\label{loesung} 
 
In this appendix an efficient numerical method for solving the master equation 
(\ref{Master_gedaechtnis}) is outlined.

First, we make a rearrangement of \Eq{Master_gedaechtnis}
splitting it into a PDM-part and an additional part 
 that disappears in the limit $\beta\to\infty$, namely 
\begin{equation}\label{Master2} 
\punkt{\rho}(t)=-i\Bigl[H_{eff}\rho(t)-\rho(t)H^{\dagger}_{eff}\Bigr]+{\bf L}\rho(t). 
\end{equation} 
Thereby we have introduced
the effective non-Hermitian Hamiltonian 
\begin{equation}\label{Heff} 
H_{eff}=H_{dr.}-ib\proj{g}{g}, 
\end{equation} 
the damping operator 
\begin{eqnarray} 
{\bf L}\rho(t)&=&2b\proj{g}{g}\rho(t)\proj{g}{g}\label{liouvilleoperator}\\ 
&+&b\int_0^t d\tau\sum_{n\not=g}\Bigl\{\proj{n}{n}\rho(t) 
\proj{g}{g}w_n(\tau)+\mbox{h.c.}\Bigr\}\nonumber 
\end{eqnarray} 
and a memory function 
\begin{equation}\label{wdef} 
w_n=
\lim_{\Omega\to\infty}\Omega e^{-\Omega \tau}-\alpha_n^0(\tau). 
\end{equation} 
In the PDM-limit, i.e. $\beta \to \infty$,
 the second term on the right side of \Eq{liouvilleoperator}
disappears and the master equation reduces to  
\Eq{PDMMaster}. 
Integration of \Eq{Master2} yields  
\begin{eqnarray}  
{}\!\!\!\!\!\!\!\!\!\!\!\!\!\!\!\!\!\!&&\rho(t)=U(t)\rho(0)U^{\dagger}(t)+\\ 
&&\qquad\qquad\ \ \int_0^\infty dt'\Theta(t-t')U(t-t') 
{\bf L}\rho(t')U^{\dagger}(t-t')\nonumber 
\end{eqnarray} 
where $U(t)=\exp[-iH_{eff}t]$ is a non unitary time evolution operator. If the initial condition 
is taken to be $\rho(0)=\proj{g}{g}$, the Laplace transformed  matrix elements of the density operator 
$\tilde{\rho}_{ij}(z)\equiv{\cal L}_z\rho_{ij}(t)\equiv 
\int_0^{\infty}\;dt e^{izt}\rho_{ij}(t)$ 
become 
\begin{eqnarray} 
&&\tilde{\rho}_{lk}(z)=H_{lggk}(z)\Bigl[1+2b\tilde{\rho}_{gg}(z)\Bigr]+\label{laplacetrafo}\\ 
&&b\sum_{n\not= g}\Bigl\{H_{lngk}(z)\tilde{w}_n(z)\tilde{\rho}_{ng}(z)+ 
H_{lgnk}(z)\tilde{w}^*_n(-z)\tilde{\rho}_{gn}(z)\Bigr\},\nonumber\\[2mm] 
&&H_{abcd}(z)=\label{Habcd}\\ 
&&\frac{1}{2\pi}\int_{-\infty}^{\infty}\!dz_1{\cal L}_{z_1+i0}\brkt{a}{U(t)}{b} 
\left[{\cal L}_{z_1-z+i0}\brkt{d}{U(t)}{c}\right]^*\nonumber 
\end{eqnarray} 
with $\tilde{w}_n(z)=1-\tilde{\alpha}_n^0(z)$. 
The Laplace transformed transition amplitudes ${\cal L}_z\brkt{i}{U(t)}{j}$ appearing in 
\Eq{Habcd} are easily calculated \cite{AlberZoller,Wir}. 
\begin{eqnarray} 
&&{\cal L}_z\brkt{g}{U(t)}{g}=\frac{i}{z+ib-\equer-\Sigma(z)},\label{agz}\\ 
&&{\cal L}_z\brkt{g}{U(t)}{n}={\cal L}_z\brkt{n}{U(t)}{g}\\ 
&&\qquad\qquad\qquad\ =\frac{-i\varepsilon_0\dn}{(z-\epsilon_n)[z+ib-\equer-\Sigma(z)]},\nonumber\\[2mm] 
&&{\cal L}_z\brkt{n}{U(t)}{l}=\label{anlz}\\ 
&&\quad\quad\frac{1}{z-\epsilon_n}\Bigl\{i\delta_{nl}+ 
\frac{i\varepsilon_0^2\dn \darg{l}}{(z-\epsilon_l)[z+ib-\equer-\Sigma(z)]}\Bigr\}\nonumber 
\end{eqnarray}  
where $\Sigma(z)=\sum_{n\not= g}\frac{|\varepsilon_0 \dn|^2}{z-\epsilon_n}$  
is the self energy of state $\ket{g}$ and the 
Kronecker-symbol $\delta_{nl}$ turns into a Dirac-delta function $\delta(\epsilon_n-\epsilon_l)$ 
for  energy normalized continuum states $\ket{\epsilon_n}$ and $\ket{\epsilon_l}$. 
For the subsequent treatment it
is convenient to introduce the expressions 
\begin{eqnarray} 
\alpha_n(z)&=&\frac{\tilde{\rho}_{ng}(z)}{[1+2b\tilde{\rho}_{gg}(z)]\varepsilon_0\dn},\label{alphadef}\\ 
\beta_n(z)&=&\frac{\tilde{\rho}_{gn}(z)}{[1+2b\tilde{\rho}_{gg}(z)]\varepsilon_0\dn},\label{betadef}\\ 
E_n(z)&=&\frac{H_{gngg}(z)}{\varepsilon_0\dn},\quad 
F_n(z)=\frac{H_{ggng}(z)}{\varepsilon_0\dn}.\label{FEdef} 
\end{eqnarray} 
Thus \Eq{laplacetrafo} and Laplace transformation of \Eq{Master2} yield 
\begin{eqnarray} 
&&\tilde{\rho}_{gg}(z)=\frac{{\cal K}(z)}{1-2b{\cal K}(z)},\label{rggz}\\ 
&&\tilde{P}_{ion}(z)=\label{pionz}\\ 
&&\quad\frac 1 {2\pi(z+i0)(1-2b{\cal K}(z))} 
\int_0^{\infty}d\epsilon\bigl[\alpha_\epsilon(z)-\beta_\epsilon(z)\bigr]\nonumber 
\end{eqnarray} 
with the characteristic kernel 
\end{multicols} 
$\!\!\!\!$\hrulefill$\quad\hfill\ $\\ 
\begin{equation}\label{kernel} 
{\cal K}(z)=H_{gggg}(z)+b\Bigl(\sum_{n\not=g}+\int_0^{\infty}dn(\epsilon)\Bigr) 
|\varepsilon_0\dn|^2\Bigl\{E_n(z)\tilde{w}_n(z)\alpha_n(z)+F_n(z)\tilde{w}_n^*(-z)\beta_n(z)\Bigr\} 
\end{equation} 
and with
\begin{eqnarray} 
\alpha_l(z)&=&\frac{\tilde{w}_l^*(-z)C_l(z)|\varepsilon_0\darg{l}|^2\Bigl(F_l(z)+S_l(z)\Bigr)+ 
\bigl(1-J_l(z)\bigr)\Bigl(E_l(z)+T_l(z)\Bigr)} 
{\bigl(1-G_l(z)\bigr)\bigl(1-J_l(z)\bigr)-(\tilde{w}_l(z)\tilde{w}_l^*(-z))^2 C_l(z)^2|\darg{l}|^4},\label{alphaz}\\ 
\beta_l(z)&=&\frac{w_l(z)C_l(z)|\varepsilon_0 \darg{l}|^2\Bigl(E_l(z)+T_l(z)\Bigr)+\bigl(1-G_l(z)\bigr)\Bigl(F_l(z)+S_l(z)\Bigr)} 
{\bigl(1-G_l(z)\bigr)\bigl(1-J_l(z)\bigr)-(\tilde{w}_l(z)\tilde{w}_l^*(-z))^2 C_l(z)^2|\darg{l}|^4}.\label{betaz} 
\end{eqnarray} 
The non-diagonal couplings  
$(\alpha_l,\beta_l\leftrightarrow \alpha_n,\beta_n,\quad n\not=l)$ 
due to the sum in \Eq{laplacetrafo} give rise to
the expressions $S_l(z)$ and $T_l(z)$ appearing in 
Eqs.(\ref{alphaz}) and (\ref{betaz}), namely 
\begin{eqnarray} 
S_l(z)&=&b\Bigl\{\sum_{n\not=\{g,l\}}+\int_0^{\infty}dn(\epsilon)\Bigr\} 
|\varepsilon_0\dn|^2\Bigl[\beta_n(z)\frac{F_n(z)-F_l(z)}{\epsilon_l-\epsilon_n} 
-\alpha_n(z)\frac{F_l(z)-E_n(z)}{\epsilon_n-\epsilon_l-z-2i0}\Bigr],\label{Svonl}\\ 
T_l(z)&=&b\Bigl\{\sum_{n\not=\{g,l\}}+\int_0^{\infty}dn(\epsilon)\Bigr\} 
|\varepsilon_0\dn|^2\Bigl[\alpha_n(z)\frac{E_n(z)-E_l(z)}{\epsilon_l-\epsilon_n} 
-\beta_n(z)\frac{E_l(z)-F_n(z)}{\epsilon_n-\epsilon_l+z+2i0}\Bigr].\label{Tvonl} 
\end{eqnarray} 
The diagonal couplings of the $\alpha_n$ and $\beta_n$ yield $J_l(z),G_l(z)$ and 
$C_l(z)$, i.e.
\begin{eqnarray} 
C_l(z)&=&b\Theta(-\epsilon_l)\frac{E_l(z)-F_l(z)}{z+2i0},\label{CL}\\ 
G_l(z)&=&\frac{b\tilde{w}_l(z)}{2\pi}\int_{-\infty}^{\infty} 
\frac{dz_1\Bigl\{1+\Theta(-\epsilon_l)|\varepsilon_0\darg{l}|^2 
\bigl[z_1+ib-\equer-\Sigma(z_1+i0)\bigr]^{-1}(z_1-\epsilon_l+i0)^{-1}\Bigr\}} 
{\bigl[z_1-z-ib-\equer-\Sigma(z_1-z-i0)\bigr](z_1-\epsilon_l+i0)}, 
\label{GlpKl}\\ 
J_l(z)&=&\frac{b\tilde{w}_l^*(-z)}{2\pi}\int_{-\infty}^{\infty} 
\frac{dz_1\Bigl\{1+\Theta(-\epsilon_l)|\varepsilon_0\darg{l}|^2 
\bigl[z_1-z-ib-\equer-\Sigma(z_1-z-i0)\bigr]^{-1}(z_1-z-\epsilon_l-i0)^{-1}\Bigr\}} 
{\bigl[z_1+ib-\equer-\Sigma(z_1+i0)\bigr](z_1-z-\epsilon_l-i0)}.\label{JlpLl} 
\end{eqnarray} 
\hfill \hrulefill\\ 
\begin{multicols}{2} 
All information about the quantum system is contained in the self energy $\Sigma(z)$. In order to solve 
Eqs.(\ref{alphaz}-\ref{Tvonl}) for a given value of $z$, the coefficients  
$H_{gggg},E_l,F_l,J_l$ and $G_l$ have to be calculated. 
Using the energy levels and dipole matrix elements  
of Eqs.(\ref{onechannel})  
the self energy becomes \cite{AlberZoller} 
\begin{eqnarray} 
\Sigma(z)&=&\delta\omega-i\frac{\gamma}{2}+i\gamma\frac{1}{1-\exp(-2\pi i\nu(z))]}\label{sigmaz}\\ 
\mbox{with}\qquad\nu(z)&=&(-2z)^{-1/2}+\alpha\label{nuvonz}  
\end{eqnarray}
and with the (non-resonant) quadratic Stark-shift contribution $\delta\omega$.
In\cite{Wir} we calculated the quantity $H_{gggg}$ ($f(z)$ in that work) by
contour integration. 
In an analogous way the quantities $E_l,F_l,J_l$ and $G_l$ 
can be calculated but for sake of brevity we do not give them here explicitly. 
Starting with $S_n=T_n=0$, Eqs.(\ref{rggz}-\ref{Tvonl}) are solved by iteration.   
Actually we found the non-diagonal 
coupling terms $S_l,T_l$ to be very small in comparison with
the diagonal couplings so that this iteration converges 
very rapidly.   
\end{multicols} 
\end{appendix} 
\begin{multicols}{2} 
 
\end{multicols} 
\newpage
\end{document}